\documentclass[twocolumn]{aastex631}
\usepackage{lineno}
\usepackage{graphicx}

\begin{document}

\title{Indication of $\gamma$-Ray Quasi-periodicity in GB6 J1037+5711 from Multi-technique Timing Analysis}

\email{malikzahoor313@gmail.com}
\author[0000-0002-8955-3212]{Zahoor Malik}
\affiliation{Department of Physics, National Institute of Technology, Srinagar 190006, India.}

\email{darprince46@gmail.com}
\author{Sikandar Akbar}
\affiliation{Department of Physics, University of Kashmir, Srinagar 190006, India.}

\email{shahzahir4@gmail.com}
\author{Zahir Shah}
\affiliation{Department of Physics, Central University of Kashmir, Ganderbal 191201, India.}

\author{Seemin Rubab}
\affiliation{Department of Physics, National Institute of Technology, Srinagar 190006, India.}

\begin{abstract}
We report the indication of a long-term quasi-periodic oscillation (QPO) in the $\gamma$-ray emission of the BL~Lac object 4FGL~J1037.7+5711 (GB6~J1037+5711) using more than 17 years of monthly binned \textit{Fermi}-LAT observations. Since blazar $\gamma$-ray variability is typically dominated by stochastic red-noise processes arising from turbulent jet activity and accretion fluctuations, we applied multiple independent timing techniques to test the presence of periodic modulation. These include the 
Lomb--Scargle periodogram (LSP), weighted wavelet $Z$-transform (WWZ), first-order autoregressive red-noise modeling (REDFIT), and epoch-folding analysis. The LSP reveals a significant periodic signal at $478.74 \pm 17.55$ days, exceeding the $99.99\%$ confidence level derived from Monte Carlo simulations. The WWZ analysis independently recovers a comparable period of $474.72 \pm 27.24$ days at a significance of $99.7\%$, while the REDFIT analysis identifies a similar periodic feature at $481.67 \pm 35.41$ days above the $99\%$ confidence level. The epoch-folding analysis further confirms the same modulation timescale. The small differences in the period estimates across methods are expected given the distinct mathematical frameworks and sensitivity functions of each technique. The long-term flux distribution of the source is better described by a lognormal profile than a Gaussian, suggesting that the underlying variability arises from multiplicative processes. The indication of a $\sim$478~day 
QPO, independently confirmed across all four timing techniques, may be associated with the orbital dynamics of a supermassive binary
black hole system driving Newtonian jet precession or periodic Doppler-factor modulation, Lense--Thirring precession of the inner 
accretion disk around a rapidly spinning SMBH, or accretion-driven instabilities at the disk--jet interface subsequently amplified through relativistic beaming.
\end{abstract}



\section{Introduction}
\label{sec:intro}

Blazars are among the most extreme subclasses of active galactic nuclei (AGNs), exhibiting rapid and large-amplitude variability across the entire electromagnetic spectrum. Their relativistic jets are closely aligned with the line of sight of observer, resulting in strong Doppler boosting effects that enhance their observed emission \citep{2018AJ....155...31H}. At the center of these systems lies a supermassive black hole (SMBH) with masses typically ranging from $10^{6}$--$10^{10} M_{\odot}$ \citep{2015A&A...576A.122E,2019MNRAS.484.5785G}. The observed radiation from blazars is predominantly non-thermal in origin \citep{1980ARA&A..18..321A,1995PASP..107..803U,2017ApJS..229...21X}. Their broadband spectral energy distribution (SED) typically shows a double-peaked structure, where the low-energy component (radio to X-rays) is generally attributed to synchrotron radiation from relativistic electrons in the jet, while the high-energy component (X-rays to $\gamma$-rays) is commonly explained through inverse Compton processes \citep{2021MNRAS.506.3791R,2021RAA....21...75R}.

Based on the strength of optical emission lines, blazars are broadly classified into two categories: flat-spectrum radio quasars (FSRQs) and BL Lacertae objects (BL Lacs). FSRQs exhibit emission line equivalent widths greater than 5 \AA, whereas BL Lac objects show weak or nearly featureless spectra with equivalent widths below this threshold \citep{1995PASP..107..803U,2017ApJS..229...21X,2018AJ....155...31H}. Depending on the location of the synchrotron peak frequency, blazars are further classified as low-synchrotron-peaked, intermediate-synchrotron-peaked, and high-synchrotron-peaked sources \citep{2010ApJ...716...30A,2017AJ....154...42H,2022NewA...9001666I}.

Variability is one of the defining observational characteristics of blazars and is observed over timescales ranging from minutes to decades. Among different forms of variability, quasi-periodic oscillations (QPOs) have attracted considerable interest because they can provide important clues about the physical processes occurring in the innermost regions of AGNs and relativistic jets. QPO signatures have been reported across multiple wavebands, from radio to $\gamma$-rays, with claimed periods spanning from minutes to years. On short timescales, minute-scale periodic variations were reported in OJ 287 \citep{1985Natur.314..148V,1985Natur.314..146C}. A 7.6-day QPO was detected in CTA 102 and interpreted as arising from helical motion of an emitting region in the jet \citep{Sarkar2020}. Similarly, \citet{Zhou2018} reported a 34.5-day QPO in the $\gamma$-ray light curve of PKS 2247$-$131, while \citet{2019MNRAS.484.5785G} identified a 71-day periodic signal in B2 1520+31. A 120-day QPO was also detected in J1359+4011 and linked to thermal instabilities in the accretion disk \citep{King2013}. Additionally, \citet{Roy2022} reported two QPOs in PKS 1510$-$089 with periods of 3.6 days and 92 days.

Long-term periodicities on year-like timescales have also been reported in several blazars. A 5.7-year QPO was detected in AO 0235+16 \citep{Raiteri2001}, while a 3-year periodicity was reported in 3C 66A \citep{OteroSantos2020}. \citet{Ren2021} identified a $4.69\pm0.14$ year QPO in PKS J2134$-$0153, and \citet{Gong2022} reported a 2.8-year periodicity in PKS 0405$-$385. In the $\gamma$-ray band, one of the earliest significant detections was reported in PG 1553+113 with a period of nearly two years \citep{Ackermann2015}. Since then, more than 30 candidate QPO signals have been identified in \textit{Fermi}-LAT blazar light curves \citep{Sandrinelli2016, 2017MNRAS.471.3036P,
Zhang2017, Bhatta2019, Zhang2021, Zhang2023}.

Several physical mechanisms have been proposed to explain QPOs in blazars. One of the most widely discussed scenarios involves binary supermassive black hole systems, where the observed periodicity reflects orbital motion of a secondary black hole around a primary SMBH \citep{Komossa2006}. The most prominent example is OJ 287, which shows a $\sim$12-year optical periodicity interpreted in terms of a binary SMBH system \citep{1988ApJ...325..628S,
Valtonen2006}. Other possible explanations include helical jet motion \citep{Camenzind1992, Mohan2015}, jet precession \citep{Liska2018}, accretion disk instabilities \citep{bhatta2020nature}, and magnetic reconnection processes in relativistic jets \citep{Huang2013}.

The BL Lac object 4FGL J1037.7+5711 (also known as GB6 J1037+5711), listed in the \textit{Fermi}-LAT 4FGL-DR2 catalog, exhibits a variability index greater than 21.67, making it an interesting candidate for investigating long-term variability. It is a relatively bright source that has shown activity extending into the very-high-energy (VHE) regime, with 10 VHE photons and 325 high-energy photons detected in total \citep{PSHIRKOV2026100492}. The VHE flare coincided with a prolonged phase of enhanced high-energy activity. A transient TeV event from this source was independently detected by the High Altitude Water Cherenkov Observatory (HAWC) during MJD 58298--58300 \citep{2018ATel11806....1W}. Despite its brightness, the source lacks a secure redshift measurement. SDSS DR7 spectroscopy reported a tentative redshift of $z=0.83\pm0.0004$ with very low confidence \citep{2009ApJS..182..543A}, while later observations revealed a featureless optical spectrum and constrained the source to $z>0.25$ \citep{2020MNRAS.497...94P,2023MNRAS.518.2675K}. Broadband SED modeling has also suggested a possible redshift of $z=1.14$ \citep{2022Univ....8..587F}.

\citet{2024A&A...689A..35C} has reported indication of optical quasi-periodic behavior in this source, detecting periods of $1296\pm580$ days using the generalized Lomb--Scargle method and 349 days using the weighted wavelet Z-transform, with significance levels greater than $2\sigma$. However, a detailed investigation of long-term $\gamma$-ray periodic behavior in this source is still lacking.

In this work, we investigate the long-term $\gamma$-ray variability of 4FGL J1037.7+5711 using more than 17 years (MJD 54728-61118) of \textit{Fermi}-LAT observations. We employ multiple independent timing techniques, including LSP, WWZ, autoregressive red-noise modeling (AR(1)), and epoch-folding analysis to search for the periodic signatures. Our analysis reveals evidence for a year-scale QPO in the $\gamma$-ray light curve of this source, which may provide important insights into the physical mechanisms driving periodic variability in blazars.

The remainder of this paper is organized as follows. In Section~\ref{sec:obs}, we describe the \textit{Fermi}-LAT data selection and reduction procedures adopted in this work. Section~\ref{sec:timing} presents the timing analysis of the $\gamma$-ray light curve of 4FGL J1037.7+5711 using the LSP, WWZ, REDFIT AR(1) modeling, and epoch-folding techniques. Section~\ref{sec:distribution} presents the flux distribution analysis of the source and investigates the statistical nature of its long-term variability. Finally, Section~\ref{sec:discussion} summarizes our main findings and discusses their possible physical implications.

\section{Observations and Data Reduction}
\label{sec:obs}
\subsection{\textit{Fermi}-LAT}

The Large Area Telescope (LAT) onboard the \textit{Fermi Gamma-ray Space Telescope} is a pair-conversion $\gamma$-ray detector designed to observe photons in the energy range from $\sim$20 MeV to beyond 500 GeV. Since its launch in 2008, \textit{Fermi}-LAT has operated primarily in all-sky survey mode, scanning the entire sky approximately every three hours due to its large field of view of about 2.4 sr \citep{2009ApJ...697.1071A}. This observing strategy makes it particularly well suited for long-term variability studies of active galactic nuclei.

For this work, we analyzed more than 17 years of \textit{Fermi}-LAT Pass 8 observations of the BL Lac object 4FGL J1037.7+5711, covering the time interval MJD 54728--61118. The analysis was performed using \texttt{Fermitools} along with the \texttt{Fermipy} package \citep{2017ICRC...35..824W}, following standard analysis procedures recommended by the \textit{Fermi} Science Support Center.

We selected SOURCE class events (\texttt{evclass=128}) and FRONT+BACK converting photons (\texttt{evtype=3}) in the energy range 0.1--100 GeV within a region of interest (ROI) of $24^{\circ}\times24^{\circ}$ centered on 4FGL J1037.7+5711. A zenith angle cut of $90^{\circ}$ was applied to reduce contamination from $\gamma$ rays originating from the Earth's limb. The analysis used a spatial bin size of $0.1^{\circ}$ and eight logarithmic energy bins per decade.

The source model included all cataloged sources within a $30^{\circ}$ region centered on the target using the \textit{Fermi}-LAT Fourth Source Catalog Data Release 2 (4FGL-DR2). The Galactic diffuse emission was modeled using \texttt{gll\_iem\_v07.fits}, while the isotropic diffuse background was represented using \texttt{iso\_P8R3\_SOURCE\_V3\_v1.txt}. We adopted the post-launch instrument response functions \texttt{P8R3\_SOURCE\_V2}.

An initial maximum-likelihood optimization of the ROI model was performed using \texttt{gtlike}. To improve the fitting procedure, the normalization parameters of all sources located within $12^{\circ}$ of the target were allowed to vary. The normalization parameters of sources with TS $<10$ were fixed. For the target source, we first allowed only the normalization parameter to vary and performed a likelihood fit, followed by a second fit in which both the normalization and spectral index parameters were left free. After obtaining the optimized source model, we generated the $\gamma$-ray light curve of 4FGL J1037.7+5711 using 30-day time bins. The resulting long-term $\gamma$-ray light curve was subsequently used for the timing analysis presented in the following sections.

\section{Quasi-periodic Oscillation Analysis}
\label{sec:timing}

To investigate the presence of possible quasi-periodic behavior in the long-term $\gamma$-ray light curve of 4FGL J1037.7+5711, we employed several independent timing analysis techniques that are widely used for unevenly sampled astronomical data. These include the LSP, WWZ, AR(1), and epoch-folding analysis. Furthermore, we performed extensive Monte Carlo simulations to assess the statistical significance of the detected periodic features and to distinguish genuine periodic signals from stochastic variability. The individual methods and their corresponding results are discussed in the following subsections.

\subsection{Lomb--Scargle Periodogram (LSP)}

The Lomb--Scargle periodogram (LSP) is one of the most commonly used techniques for detecting periodic signals in unevenly sampled astronomical time-series data \citep{lomb1976least,scargle1982studies}. Its ability to efficiently handle irregularly sampled light curves makes it particularly useful for long-term $\gamma$-ray variability studies of blazars. In this work, we employed the \texttt{astropy} implementation of the Lomb--Scargle algorithm\footnote{\url{https://docs.astropy.org/en/stable/timeseries/lombscargle.html}}, incorporating the observed flux uncertainties to obtain a more reliable estimate of the power spectrum. The frequency search range was chosen between $f_{\rm min}=1/T$ and $f_{\rm max}=1/(2\Delta T)$, where $T$ represents the total duration of the light curve and $\Delta T$ corresponds to the characteristic sampling interval \citep{vanderplas2018understanding}.

\begin{figure*}
    \centering
    \includegraphics[scale=0.9,angle=0]{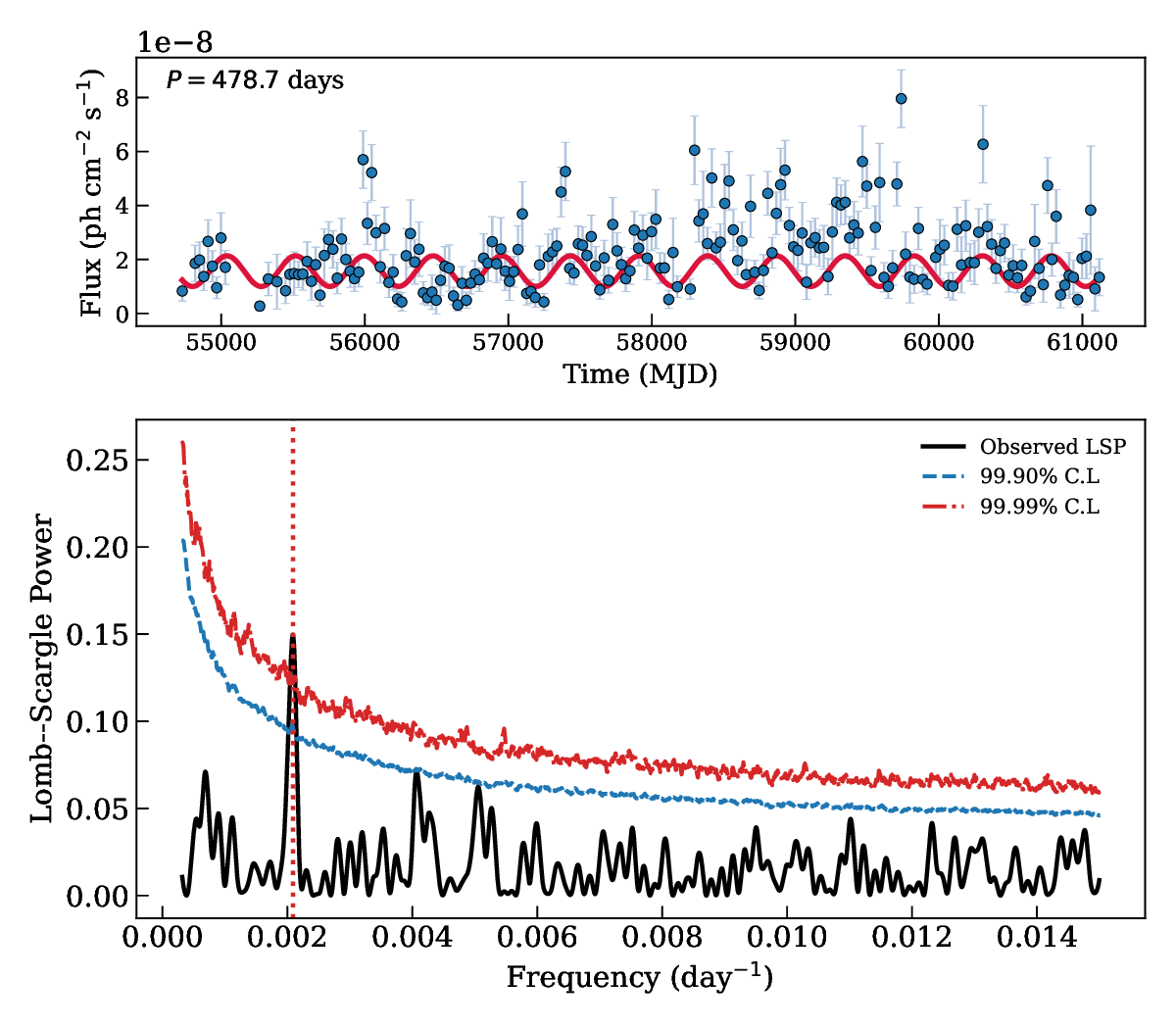}
    \caption{Top: Monthly binned $\gamma$-ray light curve of 4FGL J1037.7+5711 fitted with a sinusoidal model corresponding to the detected quasi-periodic modulation. The source exhibits a periodic timescale of $\sim478.74$ days and spans approximately $\sim13$ observed cycles over the entire \textit{Fermi}-LAT monitoring duration. Bottom: Lomb--Scargle periodogram of the monthly binned $\gamma$-ray light curve showing a prominent peak at a frequency of $0.002089$ day$^{-1}$, corresponding to a period of $\sim478.74$ days. The detected peak exceeds the 99.99\% confidence level derived from $10^{5}$ Monte Carlo simulated light curves generated using the method of \citet{emmanoulopoulos2013generating}.}
    \label{fig:lsp}
\end{figure*}

The resulting periodogram for the $\gamma$-ray light curve of 4FGL J1037.7+5711 shows a prominent peak at a frequency of $0.002089 \pm 0.000077$ day$^{-1}$, corresponding to a period of $478.74 \pm 17.55$ days. The uncertainty in the detected period was estimated by fitting a Gaussian profile to the dominant peak and adopting the half-width at half-maximum as the associated error. The resulting LSP is shown in Figure~\ref{fig:lsp}. In the upper panel of Figure~\ref{fig:lsp}, we also present the monthly binned $\gamma$-ray light curve overlaid with the sinusoidal modulation corresponding to the detected $\sim478.74$ day period, which illustrates the repeated variability pattern across multiple cycles ($\sim13$) during the entire \textit{Fermi}-LAT observational baseline. To assess the statistical reliability of the detected feature, we computed the false alarm probability (FAP) using the \texttt{LombScargle.false\_alarm\_probability()} 
routine implemented in \texttt{astropy}. This method provides an analytic estimate of the FAP based on the extreme-value statistics formalism developed by \citet{2008MNRAS.385.1279B}, accounting for the number of independent frequencies sampled by the periodogram. The dominant peak was found to exceed the $0.1\%$ significance threshold, implying that the probability of obtaining such a peak purely due to stochastic fluctuations in the light curve is less than $0.1\%$, suggesting that the detected modulation is unlikely to 
arise from random noise alone. We note, however, that the \citet{2008MNRAS.385.1279B} method assumes white Gaussian noise and may underestimate the false alarm probability in the presence of correlated red-noise variability, which typically dominates blazar light 
curves. This analytic FAP is therefore reported as an initial diagnostic only.

To further assess whether the detected signal could arise from stochastic variability, we performed extensive Monte Carlo simulations using the \texttt{DELCgen} package\footnote{\url{https://github.com/samconnolly/DELightcurveSimulation}} following the method of \citet{emmanoulopoulos2013generating}. A total of $10^{5}$ synthetic light curves were generated to reproduce both the power spectral density (PSD) and probability density function (PDF) of the observed $\gamma$-ray light curve. For each simulated light curve, LSP was computed and used to construct confidence levels across the sampled frequency range. We find that the observed peak at $\sim 479$ days exceeds the 99.99\% significance level derived from the simulations, strongly supporting the presence of a quasi-periodic signature in the $\gamma$-ray emission of 4FGL J1037.7+5711. The corresponding LSP and simulation significance results are presented in Figure~\ref{fig:lsp}.

\subsection{Weighted Wavelet Z-transform (WWZ)}

Although the Lomb--Scargle periodogram is effective in identifying periodic signals in unevenly sampled light curves, it does not provide information regarding the temporal evolution of such signals. To examine the persistence of the detected modulation over the entire observational duration, we applied the WWZ technique \citep{foster1996wavelets}. The WWZ method is particularly useful for non-stationary time series because it simultaneously analyzes the data in both time and frequency domains, allowing the evolution of periodic signals to be tracked over time.

In the WWZ formalism, the observed light curve is convolved with a localized oscillatory kernel given by

\begin{equation}
f[\omega(t-\tau)] = \exp \left[i\omega(t-\tau)-c\omega^{2}(t-\tau)^{2}\right],
\end{equation}

where $\omega$ represents the angular frequency, $\tau$ denotes the temporal translation parameter, and $c$ controls the width of the wavelet window. In this work, we adopted $c = 0.001$, which provides a suitable balance between time and frequency resolution for detecting long-term periodic features in the $\gamma$-ray light curve.

\begin{equation}
W[\omega,\tau:x(t)] = \omega^{1/2}\int x(t) f^{*}[\omega(t-\tau)]dt,
\end{equation}

where $f^{*}$ denotes the complex conjugate of the wavelet function. This formulation enables the identification of both transient and persistent periodic components in unevenly sampled data. A stable periodic signal typically appears as a concentrated region of enhanced power over an extended time interval in the time-frequency plane.

The analysis was performed using a publicly available Python implementation of the WWZ algorithm\footnote{\url{https://github.com/eaydin/WWZ}}. The resulting WWZ power spectrum of the monthly binned $\gamma$-ray light curve of 4FGL J1037.7+5711 reveals a prominent peak at a frequency of $0.002106 \pm 0.000121$ day$^{-1}$, corresponding to a period of $474.72 \pm 27.24$ days. The uncertainty associated with the detected period was estimated by fitting a Gaussian profile to the dominant peak in the time-averaged WWZ spectrum. This period is comparable with the value obtained from the Lomb--Scargle analysis, providing independent support for the presence of a year-scale quasi-periodic modulation in the source.

\begin{figure*}
    \centering
    \includegraphics[scale=0.5,angle=0]{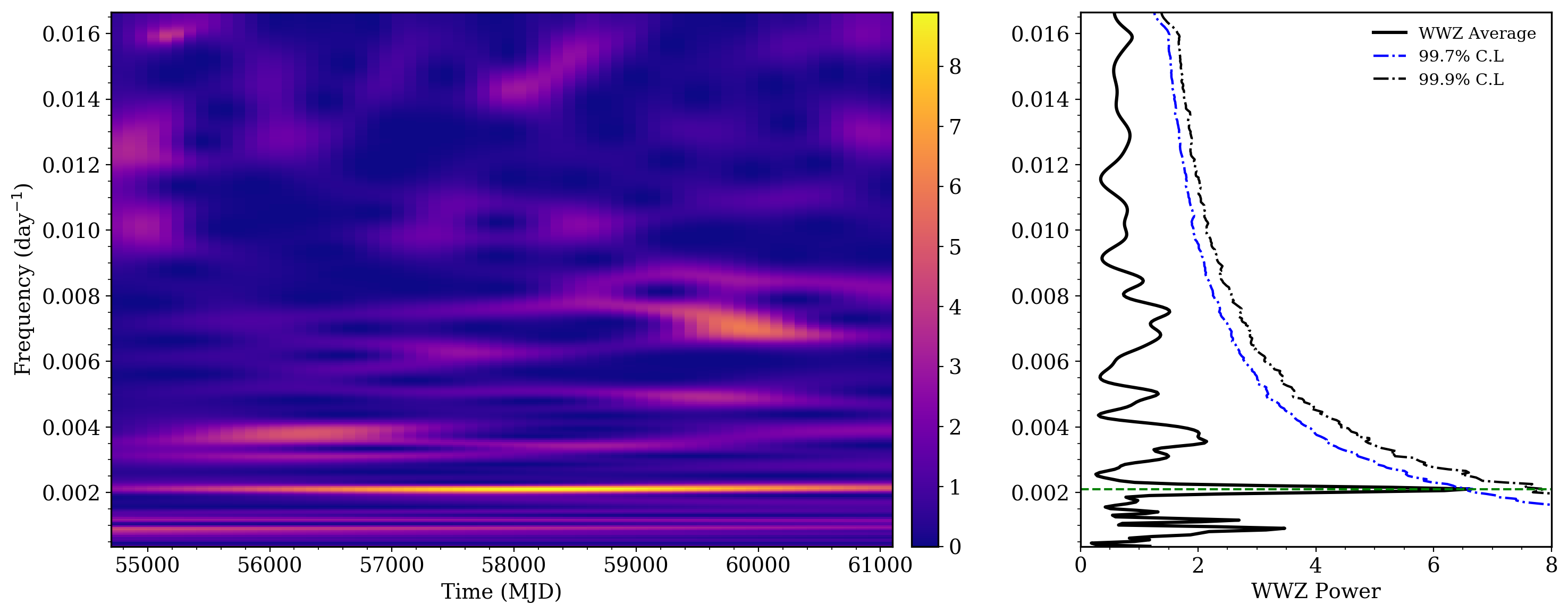}
    \caption{Left: WWZ map of the $\gamma$-ray light curve showing the evolution of power as a function of time (MJD) and frequency. Right: The time-averaged WWZ power spectrum with confidence levels derived from $5 \times 10^{4}$ Monte Carlo simulations. The blue and black dashed lines represents the 99.7$\%$ and 99.9$\%$ confidence levels respectively. The dotted green line marks the most prominent peak at a frequency of $0.002106$ day$^{-1}$, corresponding to a period of $\sim 474.72$ days.}
    \label{fig:wwz}
\end{figure*}

To further examine the significance of the detected WWZ feature, we applied the same Monte Carlo simulation framework described in Section~3.1. Using $5 \times 10^{4}$ simulated light curves generated with the \texttt{DELCgen} package following the method of \citet{emmanoulopoulos2013generating}, we estimated the significance of the observed WWZ peak. The detected feature was found to exceed the 99.7\% confidence level, further strengthening the evidence for a persistent quasi-periodic signal in the $\gamma$-ray light curve of 4FGL J1037.7+5711. The two-dimensional WWZ power map and the corresponding time-averaged WWZ spectrum are presented in Figure~\ref{fig:wwz}.

\subsection{First-order Autoregressive Process (AR(1)/REDFIT)}

Long-term $\gamma$-ray light curves of blazars are often dominated by stochastic red-noise variability arising from physical processes in the relativistic jet or accretion flow. Such variability can frequently be described using a first-order autoregressive [AR(1)] process, where the flux at a given time depends on its immediately preceding value \citep{schulz2002redfit}. The AR(1) process can be expressed as

\begin{equation}
r(t_i)=A\,r(t_{i-1})+\epsilon(t_i),
\end{equation}

where $A=\exp[-(t_i-t_{i-1})/\tau]$ represents the autoregressive coefficient associated with the characteristic timescale $\tau$, and $\epsilon(t_i)$ denotes a random driving term. The corresponding theoretical red-noise spectrum is given by

\begin{equation}
G_{rr}(f_i)=G_{0}\left(\frac{1-A^{2}}
{1-2A\cos(\pi f_i/f_{\rm Nyq})+A^{2}}\right),
\end{equation}

where $G_{0}$ is the average spectral power, $f_i$ denotes the sampled temporal frequencies, and $f_{\rm Nyq}$ represents the Nyquist frequency.

To account for the effects of red-noise contamination in the observed light curve, we applied the REDFIT method\footnote{\url{https://rdrr.io/cran/dplR/man/redfit.html}} to the monthly binned $\gamma$-ray light curve of 4FGL J1037.7+5711. This technique estimates the red-noise background directly from unevenly sampled data and enables the identification of significant periodic features above the stochastic noise component.

\begin{figure*}
    \centering
    \includegraphics[scale=0.7,angle=0]{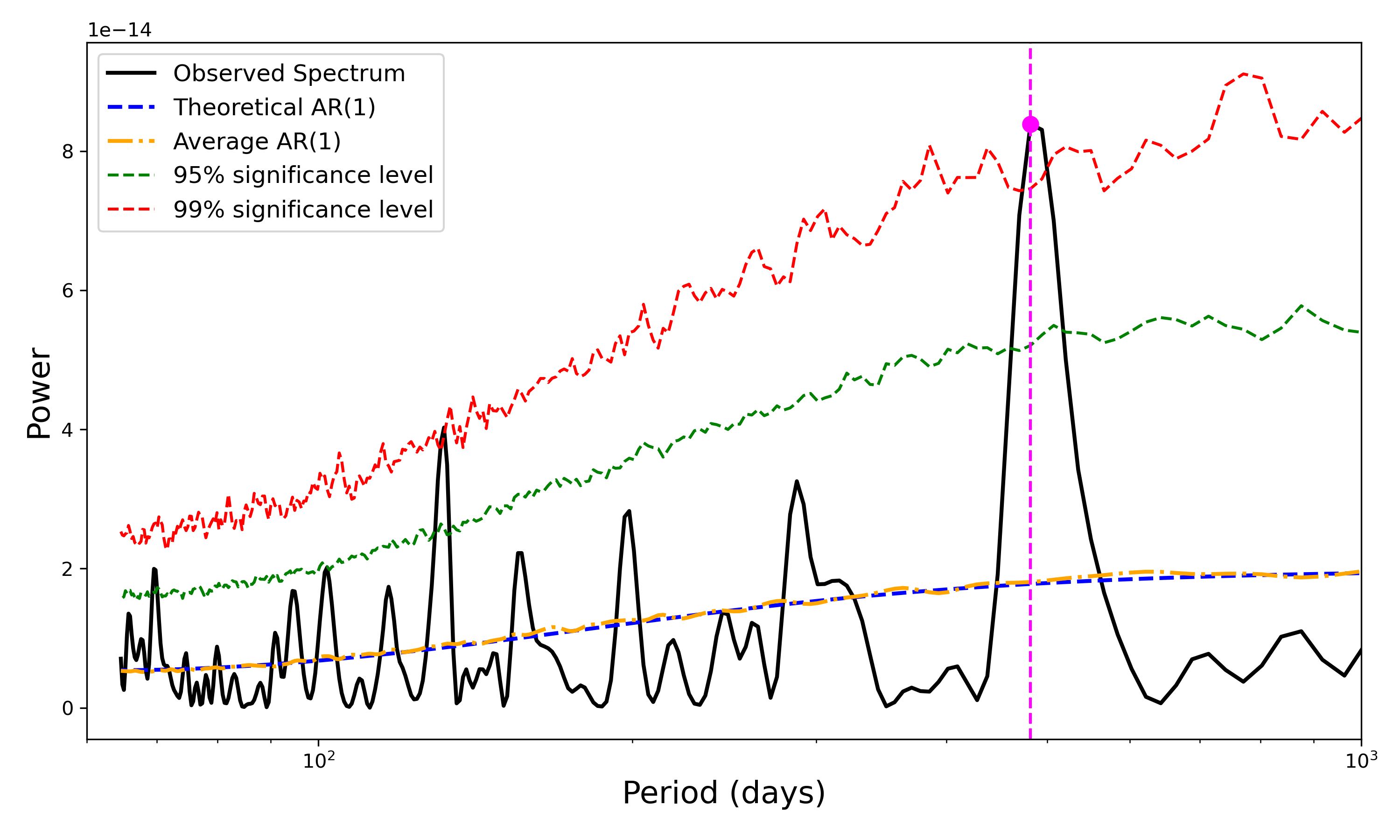}
    \caption{Red-noise-corrected REDFIT power spectrum of the $\gamma$-ray light curve of 4FGL J1037.7+5711. The black curve represents the observed spectrum, while the blue and orange curves correspond to the theoretical and ensemble-averaged AR(1) spectra, respectively. The green and red curves indicate the 95\% and 99\% Monte Carlo confidence levels. A prominent peak is detected at a frequency of $0.002076$ day$^{-1}$, corresponding to a period of $\sim 481.67$ days, which exceeds the 99\% significance threshold.}
    \label{fig:redfit}
\end{figure*}

The resulting REDFIT spectrum reveals a prominent peak at a frequency of $0.002076 \pm 0.000153$ day$^{-1}$, corresponding to a period of $481.67 \pm 35.41$ days. The uncertainty in the detected period was estimated by fitting a Gaussian profile to the dominant peak and adopting the half-width at half-maximum as the associated error estimate. The detected feature exceeds the 99\% confidence level, further supporting the presence of a year-scale quasi-periodic modulation in the $\gamma$-ray emission of 4FGL J1037.7+5711. The observed REDFIT spectrum, theoretical AR(1) spectrum, and corresponding confidence levels are shown in Figure~\ref{fig:redfit}.

\subsection{Epoch Folding}

Epoch folding is a well-established time series analysis technique, originally developed by \citet{1983ApJ...272..256L} and subsequently refined by various researchers \citep[e.g.,][]{1990MNRAS.244...93D, 1991MNRAS.251P..64D}. Unlike conventional discrete Fourier periodogram approaches, which assume that periodic signals take a sinusoidal form, epoch folding is more flexible in that it can handle periodic components of arbitrary shape. It is also insensitive to irregular data sampling, making it particularly suitable for datasets with observational gaps.

The core idea of the method is to take a time series of $N$ data points, fold 
it over a candidate (trial) period, and bin the resulting data into $M$ phase 
bins. A statistical quantity $\chi^2$ is then computed for each trial period:

\begin{equation}
    \chi^2 = \sum_{i=1}^{M} \frac{(x_i - \bar{x})^2}{\sigma_i^2},
    \label{eq:chi2}
\end{equation}

\noindent where $x_i$ is the data value in the $i$-th bin, $\bar{x}$ is the 
mean of the data, and $\sigma_i$ is the associated uncertainty.

For data consisting purely of Gaussian noise with standard deviation $\sigma_i$, 
the $\chi^2$ value is expected to be approximately equal to $M$. However, when 
the data contain a genuine periodic signal, the $\chi^2$ value deviates 
substantially from $M$, reaching a maximum at or near the true period (for details refer to \citep{1996A&AS..117..197L}). This behavior makes epoch folding a reliable tool for periodicity searches in blazar light curves \citep[e.g.,][]{2014RAA....14..933Z}.

\begin{figure*}
    \centering
    \includegraphics[scale=0.5,angle=0]{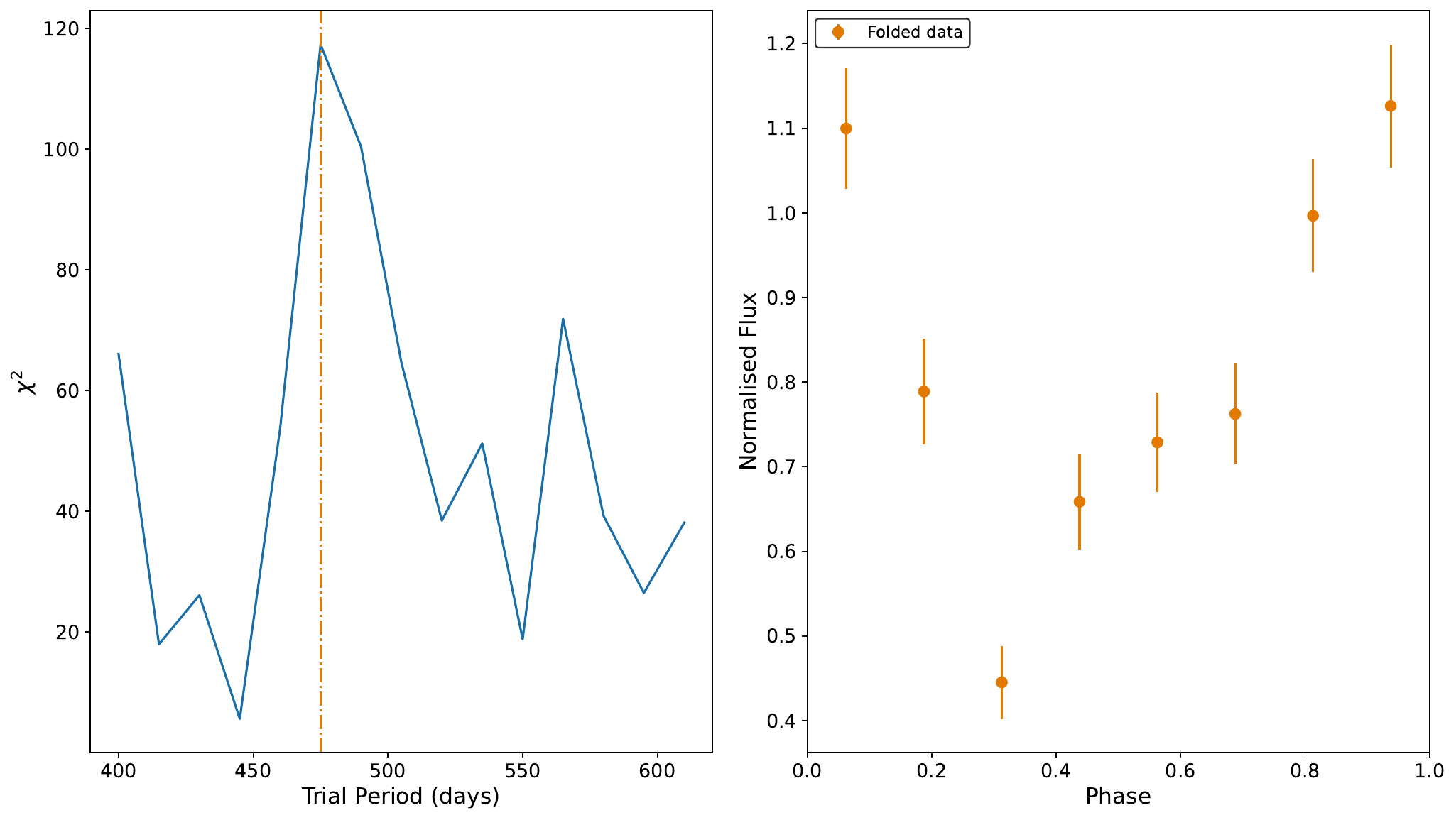}
    \caption{Epoch folding analysis of the $\gamma$-ray light curve of the blazar 4FGL J1037.7+5711.
\textit{Left panel:} The $\chi^2$ statistic computed over trial periods ranging 
from 400 to 600 days, with the peak indicating the most probable periodicity. 
\textit{Right panel:} The folded pulse profile corresponding to the best-fit 
period of $\sim$475 days, at which the $\chi^2$ value reaches its maximum.}
    \label{fig:epoch_folding}
\end{figure*}

In the present analysis, $\chi^2$ values were evaluated for trial periods 
ranging from 400 to 600\,days, using a step size equal to 15 days. Pulse profiles 
corresponding to each trial period were generated and subsequently tested for 
$\chi^2$ constancy using Equation~\ref{eq:chi2}. The left panel of 
Figure~\ref{fig:epoch_folding} displays the distribution of $\chi^2$ values 
across all trial periods considered. A prominent peak is observed at 
approximately $475$\,days, where the $\chi^2$ statistic reaches 
its maximum value of $\sim$117.383, identifying this as the most probable 
quasi-periodic timescale. The corresponding folded pulse profile, shown in 
the right panel of Figure~\ref{fig:epoch_folding}, exhibits clear flux 
modulation, further confirming the presence of periodic variability in the $\gamma$-ray light curve of 4FGL J1037.7+5711.


\section{Flux Distribution}
\label{sec:distribution}

While the timing analyses presented in Section~\ref{sec:timing} provide strong evidence for a year-scale quasi-periodic modulation in the $\gamma$-ray light curve of 4FGL J1037.7+5711, understanding the statistical nature of the underlying variability is equally important for interpreting its physical origin. Blazar variability is often governed by stochastic processes associated with particle acceleration, accretion-flow fluctuations, magnetic reconnection, or jet turbulence, and these processes can leave characteristic signatures in the flux distribution. Analysing the flux distribution of astrophysical systems therefore provides an important diagnostic for identifying the underlying physical mechanisms driving variability. A normal (Gaussian) flux distribution is generally associated with additive processes, whereas a lognormal distribution suggests multiplicative variability processes. In compact accreting black hole systems and several blazars, the observed flux distribution is often found to be lognormal. To investigate whether the long-term variability of 4FGL J1037.7+5711 follows a similar behaviour, we characterize the $\gamma$-ray flux distribution using two complementary approaches: the Anderson--Darling (AD) test and histogram fitting.

The AD test yields a test statistic of 4.13, substantially exceeding 
the critical value of 0.76 at the 5\% significance level, thereby 
rejecting the hypothesis of normality. In contrast, when applied to the 
logarithm of the flux, the AD statistic reduces to 0.65, falling below the 
critical value of 0.76 at the 5\% significance level. This indicates that the null 
hypothesis of lognormality cannot be rejected, lending statistical support 
to a lognormal flux distribution. This is further supported by the 
log-space skewness of $-0.44$, close to zero, compared to the 
flux-space skewness of 1.22, indicating a more symmetric distribution 
in log-space.

\begin{figure}
    \centering
    \includegraphics[width=0.4\textwidth]{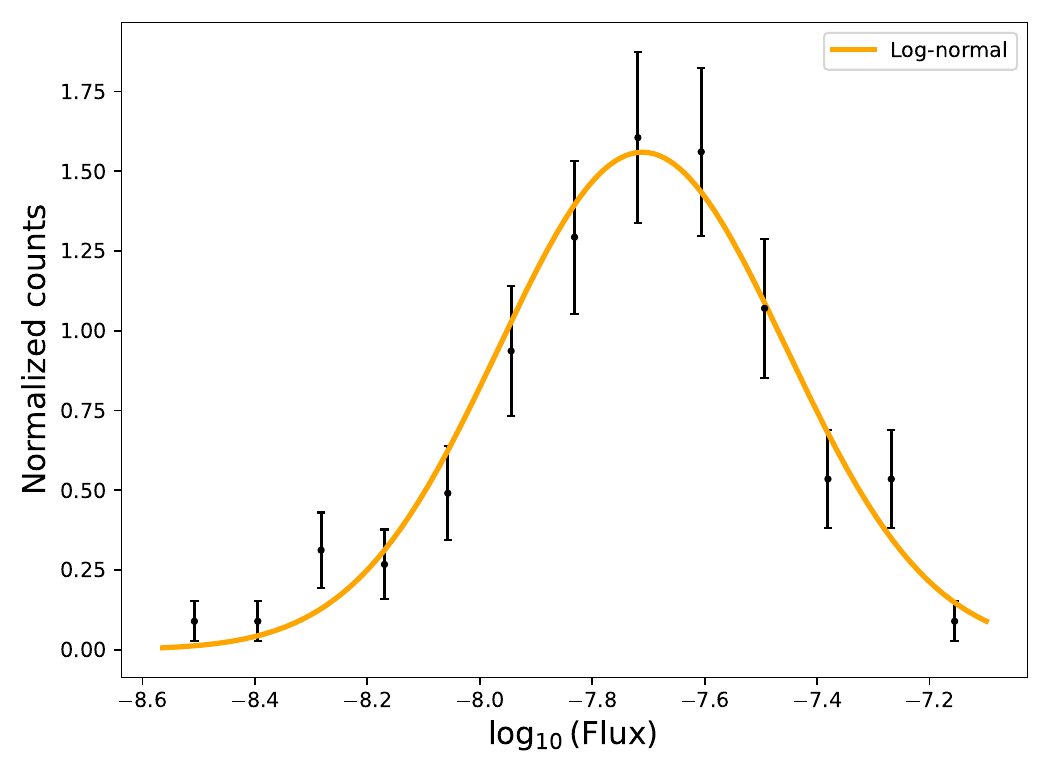}
    \caption{Normalised histogram of the $\log_{10}$ flux distribution 
    of  4FGL J1037.7+5711. The orange curve represents the best-fit 
    single log-normal function with parameters $\sigma_l = 0.256 \pm 0.015$ 
    and $\mu_l = -7.711 \pm 0.018$, yielding a reduced 
    $\chi^2 = 0.92$.}
    \label{hist_fit}
\end{figure}

We further examined the probability density function (PDF) of the flux distribution by constructing a normalized histogram of the logarithm of flux. The resulting 
histogram, shown in Figure~\ref{hist_fit}, is fitted with a Gaussian 
function in log-space:

\begin{equation}\label{eq:ln}
L(x) = \frac{1}{\sqrt{2\pi}\,\sigma_l}\,
        e^{-(x-\mu_l)^2/2\sigma_l^2}
\end{equation}

\noindent where $\mu_l$ and $\sigma_l$ denote the mean and standard 
deviation of the logarithmic flux distribution. The single log-normal 
fit yields $\sigma_l = 0.256 \pm 0.015$ and 
$\mu_l = -7.711 \pm 0.018$, with a reduced $\chi^2 = 0.92$, 
indicating a good fit to the data.

\section{Summary and Discussion}

\label{sec:discussion}
We carried out a detailed search for long-term quasi-periodic oscillations in the monthly binned \textit{Fermi}-LAT $\gamma$-ray light curve of the BL Lac object 4FGL J1037.7+5711 (GB6 J1037+5711) using more than 17 years of observations. Since blazar light curves are typically dominated by stochastic red-noise variability, multiple independent timing techniques were employed to verify the presence of any periodic signature. Our principal results are summarized as follows:

\begin{itemize}

\item The LSP analysis reveals a prominent peak at a frequency of $0.002089 \pm 0.000077$ day$^{-1}$, corresponding to a period of $478.74 \pm 17.55$ days. The detected signal exceeds the $99.99\%$ confidence level obtained from $10^{5}$ Monte Carlo simulations generated using the method of \citet{emmanoulopoulos2013generating}.

\item The sinusoidal fit to the long-term $\gamma$-ray light curve suggests that the source completes approximately $\sim13$ cycles over the entire \textit{Fermi}-LAT observational baseline, strengthening the case for a persistent year-scale quasi-periodic modulation.

\item The WWZ analysis independently recovers a consistent periodic signal at $474.72 \pm 27.24$ days ($0.002106 \pm 0.000121$ day$^{-1}$) and shows that the periodic modulation persists over a substantial fraction of the observational baseline. The detected feature reaches the $99.7\%$ significance level based on $5\times10^{4}$ simulated light curves.

\item The REDFIT analysis, which explicitly accounts for red-noise variability through an AR(1) process, identifies a dominant peak at $481.67 \pm 35.41$ days ($0.002076 \pm 0.000153$ day$^{-1}$), exceeding the $99\%$ confidence level.

\item The epoch-folding analysis independently confirms the presence of the same characteristic timescale, providing additional support for the detected modulation.

\item The long-term flux distribution of the source is better described by a lognormal profile than a Gaussian distribution, indicating that the variability may arise from multiplicative processes operating within the accretion disk--jet system.

\end{itemize}

The important outcome of this work is the consistency of 
the detected periodicity across all four independent timing 
techniques. The LSP, WWZ, REDFIT, and epoch-folding analyses 
all converge toward a characteristic timescale of $\sim$478 
days, strongly suggesting that the observed modulation is 
unlikely to be produced by stochastic red-noise fluctuations 
alone. Notably, the $\sim$478~day QPO spans approximately 13 
complete cycles over the 17-year \textit{Fermi}-LAT baseline, 
lending additional support to the persistence and physical 
reality of this modulation.

This source becomes particularly interesting when viewed in 
the context of previous studies. \citet{2024A&A...689A..35C} 
reported evidence for optical quasi-periodic behavior in this 
source, identifying periods of $1296 \pm 580$ days using the 
generalized Lomb--Scargle (GLS) method and $\sim$349 days 
using the weighted wavelet $Z$-transform. Although the optical 
periods are not identical to the $\gamma$-ray periodicity 
reported here, the presence of quasi-periodic signatures in 
multiple energy bands may indicate that the underlying 
modulation mechanism influences multiple emission zones.

Long-timescale QPOs in blazars are often interpreted through 
geometric models involving jet precession or periodic changes 
in the viewing angle of emitting regions 
\citep[e.g.,][]{rieger2004geometrical, villata1999helical, 
OteroSantos2020}. In such scenarios, even 
small variations in the jet orientation can produce 
substantial flux changes because of Doppler boosting effects. 
We note that single-jet helical motion is generally expected 
to produce QPO timescales of only $\sim$1--130~days 
\citep{rieger2004geometrical, Mohan2015}, and is 
therefore unlikely to account for the $\sim$478~day 
modulation detected here as a standalone mechanism. 
Nevertheless, jet precession driven by binary dynamics or 
periodic Doppler-factor modulation due to viewing-angle 
variations remain physically plausible interpretations for 
year-like periodicities \citep{rieger2004geometrical}. Additionally, Lense--Thirring precession 
of the inner accretion disk around a rapidly spinning SMBH 
can produce jet precession on year-like timescales 
\citep{2007Ap&SS.309..271R}, 
providing a further geometric interpretation for the 
$\sim$478~day modulation. This mechanism operates at the 
jet base on timescales of months to years depending on 
black hole spin, mass, and the characteristic disk radius 
at which precession occurs. Given the BL~Lac nature of 
4FGL~J1037.7+5711 and its strong jet-dominated emission, 
geometric interpretations of this kind remain viable.

Another possibility involves binary SMBBH systems, where the orbital dynamics of a secondary 
black hole may drive Newtonian jet precession or induce 
periodic perturbations in the accretion flow or jet 
structure \citep[e.g.,][]{1988ApJ...325..628S, 
2008Natur.452..851V, Gong2022, 
gong2024detection}. In particular, 
Newtonian-driven jet precession in a close SMBBH system 
has been shown to be well associated with observed QPO 
timescales $\gtrsim$1~yr \citep{rieger2004geometrical}, making this one of the most compelling 
scenarios for the $\sim$478~day periodicity reported here. 
Year-like periodicities observed in several blazars have 
frequently been linked to such SMBBH-driven scenarios 
\citep{Gong2022, gong2024detection}.

Intrinsic disk-driven mechanisms may also contribute. 
Accretion disk instabilities, magnetic reconnection events, 
or quasi-periodic plasma injection into the jet can generate 
long-term modulations that are subsequently amplified 
through relativistic beaming \citep[e.g.,][]{gupta2008periodic, 
bhatta2016detection, tavani2018blazar, Sandrinelli2016}. 
Such processes are well established in the literature as 
viable drivers of year-like QPOs in blazars 
\citep{Sandrinelli2016, tavani2018blazar}. As shown in Section~\ref{sec:distribution}, the long-term flux distribution 
of 4FGL~J1037.7+5711 is better described by a lognormal 
profile than a Gaussian. Such lognormal distributions 
are commonly observed in blazars and AGNs across multiple 
wavelength bands, where they are generally attributed to 
multiplicative rather than additive variability processes 
\citep{uttley2005non, shah2018log, 
10.1093/mnrasl/sly136, 10.1093/mnras/stz3108}, 
arising from physical processes within the accretion 
disc or jet \citep{Uttley_2001,2024ApJ...977..111A, 2025MNRAS.539.2185M}. When the disc and jet variability 
are coupled, fluctuations originating in the disc can 
propagate into the jet and imprint a log-normal signature 
on the observed flux distribution \citep{2024MNRAS.527.2672S}. 
The lognormal flux distribution obtained in this work 
therefore provides further support for disk-driven 
physical processes contributing to the observed 
modulation in 4FGL~J1037.7+5711.

Despite the high statistical significance obtained from 
multiple red-noise-aware tests, caution remains necessary 
because finite observing baselines can sometimes produce 
apparent periodicities in stochastic light curves. 
Continued \textit{Fermi}-LAT monitoring will therefore 
be crucial to test whether the $\sim$478~day modulation 
persists in future observations.
Future multiwavelength campaigns involving optical 
monitoring, radio VLBI observations, and broadband SED 
modeling may provide critical constraints on the physical 
origin of this QPO. In particular, periodic changes in 
jet morphology, polarization behavior, or spectral 
evolution could help distinguish between geometric, 
binary-driven, and accretion-related scenarios.

Overall, our results establish 4FGL~J1037.7+5711 as a 
compelling new candidate in the growing population of 
blazars exhibiting long-timescale $\gamma$-ray QPOs and 
highlight its importance for understanding the origin of 
periodic variability in relativistic jets.

\section{acknowledgments}
ZM acknowledges the financial support provided by the Science and Engineering Research Board (SERB), Government of India, under the National Postdoctoral Fellowship (NPDF), Fellowship reference no. PDF/2023/002995. ZM and SA express  gratitude to the Inter-University Centre for Astronomy and Astrophysics (IUCAA) in Pune, India, for the support and facilities provided. ZS is supported by the Department of Science and Technology, Govt. of India, under the INSPIRE Faculty grant (DST/INSPIRE/04/2020/002319).


%

\vspace{5mm}







\bibliography{sample631}{}
\bibliographystyle{aasjournal}


\end{document}